\documentclass{Interspeech2024}

\interspeechcameraready 

\usepackage{times}
\usepackage{soul}
\usepackage{url}
\usepackage[utf8]{inputenc}
\usepackage{graphicx}
\usepackage{amsmath}
\usepackage{amsthm}
\usepackage{booktabs}
\usepackage{algorithm}
\usepackage{algorithmic}
\usepackage[switch]{lineno}
\usepackage{amssymb}
\usepackage{stfloats}
\usepackage{tabu}                     
\usepackage{multirow}                 
\usepackage{multicol}                 
\usepackage{multirow}                
\usepackage{float}                    
\usepackage{makecell}                 
\usepackage{ulem}
\usepackage{color}  
\usepackage[cmyk]{xcolor}

\title{A New Perspective on Speaker Verification: Joint Modeling with DFSMN and Transformer}

\name[affiliation={1}]{Hongyu}{Wang}
\name[affiliation={1}]{Hui}{Li}
\name[affiliation={*}]{Bo}{Li}

\address{
  Harbin Engineering University,China
 }
\email{hongyuwang0414@gmail.com}


\begin{document}

\maketitle

\begin{abstract}
    
Speaker verification is to judge the similarity between two unknown voices in an open set, where the ideal speaker embedding should be able to condense discriminant information into a compact utterance-level representation that has small intra-speaker distances and large inter-speaker distances. We propose Voice Transformer (VOT), a novel model for speaker verification, which integrates parallel transformers at multiple scales. A deep feedforward sequential memory network (DFSMN) is incorporated into the attention part of these transformers to increase feature granularity. The attentive statistics pooling layer is added to focus on important frames and form utterance-level features. We propose Additive Angular Margin Focal Loss (AAMF) to address the hard samples problem. We evaluate the proposed approach on the VoxCeleb1 and CN-Celeb2 datasets, demonstrating that VOT surpasses most mainstream models. The code is available on GitHub\footnote{\url{https://github.com/luckyerr/Voice-Transformer_Speaker-Verification}}.
\end{abstract}

\section{Introduction}

Speaker verification is a crucial task aimed at confirming the identity of a speaker. In a typical speaker verification system, with speaker embeddings extracted from two speech segments, the system can automatically determine whether these segments belong to the same speaker or not.\enspace Typically, such a system consists of two main components.\enspace The first component is the embedding extractor [1,2,3], which extracts fixed-length representations of speakers from variable-length speech segments. The second component is the backend model [4,5,6,7], designed to compute the similarity between two speaker embedding vectors.\enspace This paper focuses on how to better extract the feature information of a speaker's voice so that the accuracy of speaker verification systems can be improved.

In recent years, Time Delay Neural Networks (TDNNs) [8] have made significant advancements in the field of speaker verification (SV) by facilitating the capture of temporal dependencies within speech features.\enspace SV techniques based on TDNNs, such as x-vector [9], have gained popularity.\enspace In particular, ECAPA-TDNN [10] has enhanced the TDNN-based architecture's capability in capturing speaker information by introducing Squeeze-and-Excitation (SE) attention [11] and multi-layer feature aggregation structures.\enspace However, TDNN-based models [9,10,12,13] require a large number of filters to extract speaker features within specific frequency regions.
Given the efficiency and popularity of Convolutional Neural Networks (CNNs)  
 [14] in the image processing field, CNN-based SV systems [15,16,17] have been extensively studied. Traditional CNN structures were combined with the Vision Transformer [18] for better extracting speaker embedding vectors.\enspace MFA-Conformer [17] combines transformer [19] and CNN modules to enhance its information extraction capabilities. Nevertheless, it still faces challenges such as lack of generalization ability and dealing with a large number of model parameters.

In this paper, to address the issue of insufficient granularity in speech features caused by the limited local modeling capability of the transformer [19] and the drawback of the fixed-scale multi-head attention mechanism, we propose the Voice-Transformer (VOT) model.\enspace The VOT model aggregates features by employing parallel transformers at multiple scales to increase feature granularity.\enspace  Additionally, we integrate Deep Feed-forward Sequential Memory Networks (DFSMNs) [20] into the self-attention mechanism of the transformer to capture local dependencies.\enspace  Furthermore, the attentive statistics pooling layer is added to focus on important frames and form utterance-level features. To tackle sample noise and boundary samples problems, we were also inspired by the loss function [22,23] to convert the traditional distance loss into angular loss and dynamically adjust the model's attention to different samples.\enspace These allow the effective extraction of both global and local features from speech to improve the performance of the VOT even under text-independent, short-speech, cross-lingual, and unconstrained conditions.

\section{The proposed method}
In this section, we introduce VOT to address the issue of insufficient granularity caused by Transformer-based models and the lack of generalization ability caused by CNN-based models.\enspace The proposed VOT combines transformers with memory models,
so that the VOT takes advantage of Transformer-based networks in modeling long-term dependencies and memory models in modeling local features.\enspace To deal with hard samples, we proposed an AAMF loss function which can increase the model's attentions to these hard samples.

\subsection{Architecture}
The proposed Voice-Transformer model is based on the Vision Transformer and is illustrated in Figure 1.
The proposed VOT model consists of three components, which are feature extractor, voice encoder and attentive statistics pooling layer [21], respectively. 
The feature extractor extracts discriminative features from voice segments. 
We use Mel filters and one-dimensional convolutions to capture the spectral characteristics of the speech.\enspace The voice encoder effectively captures the larger context and granular details of speech. By modeling both the local and global aspects of the speech, the voice encoder ensures a comprehensive representation.\enspace The attentive statistics pooling layer focuses on enhancing the representation of speech by analyzing the statistical properties of the feature.
\vspace{-0.3cm}
\begin{figure}[ht]
    \centering
    \includegraphics[width=1\linewidth]{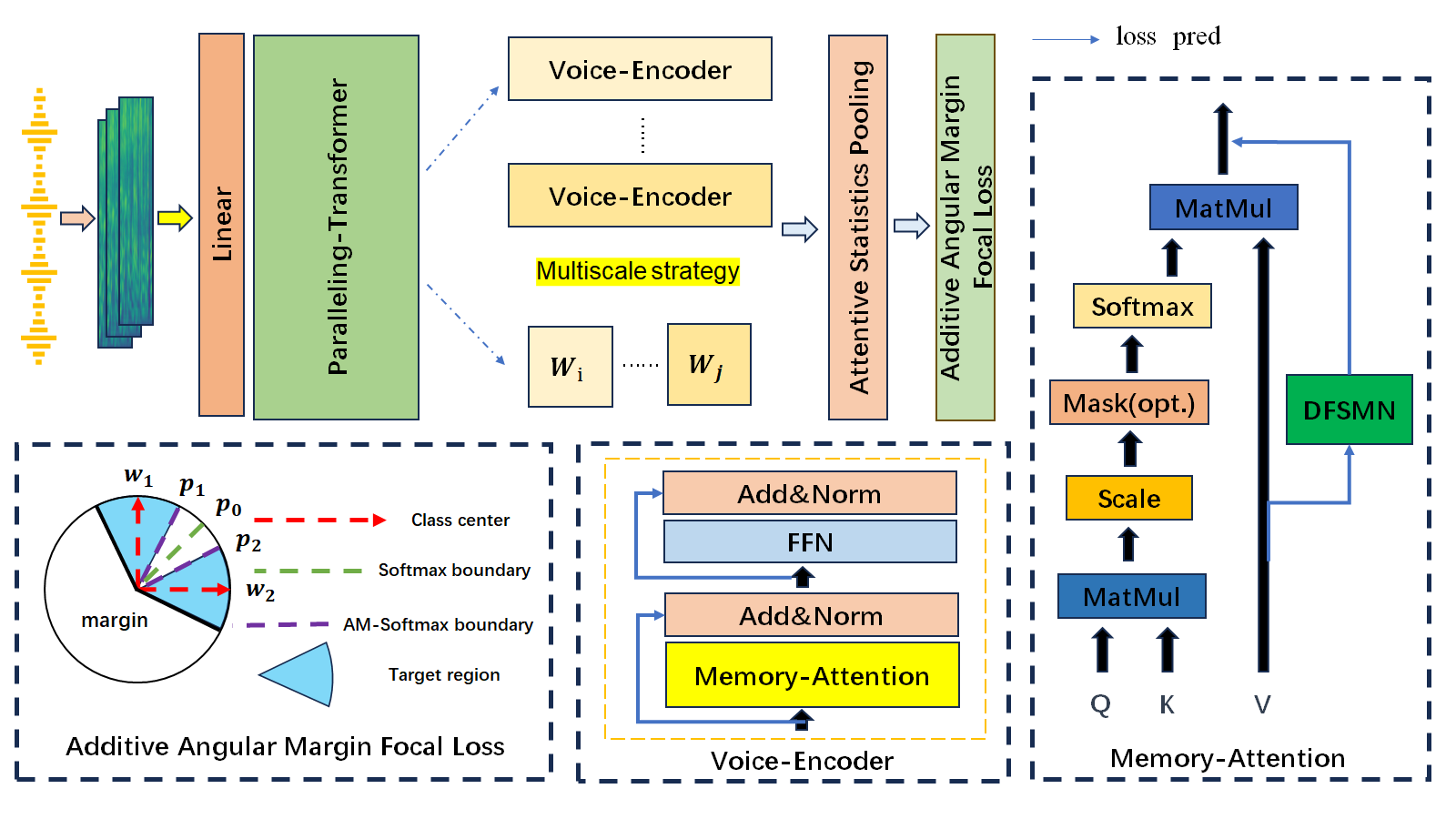}
    \caption{The Voice-Transformer architecture}
    \label{fig:1}
\end{figure}
\vspace{-0.5cm}

\subsection{Voice Encoder}
Based on transformer encoder [19], we designed a novel voice encoder that incorporates memory mechanism to enhance the transformer model's ability to model local contextual information in speech. 
As shown in Figure 1, the proposed voice encoder consists of memory-attention module, Feed-Forward Network (FFN) layers, layer normalization and residual connection structure. 

The structure of memory-attention is shown in Figure 1, which is based conventional attention mechanism.\enspace In the memory-attention, we use a Deep Feedforward Sequential Memory Network (DFSMN) [20] to process the value in the conventional attention mechanism and add the output of DFSMN to the attention result as a final output.\enspace Let $Q$, $K$, $V$ denote the query, key and value used in the memory-attention, the output of the memry-attention is:
 \vspace{-0.2cm}

\begin{equation}
\begin{aligned}
  MemorryAttention\left ( Q,K,V \right )=\\ softmax\left ( \frac{QK^{T}}{\sqrt{d_{k}}} \right )V+Memory\left ( V \right ),
\end{aligned}
\end{equation}
where $d_{k}$ is the dimension of either $Q$ or $K$.\enspace $Memory(V)$ denotes the output of DFSMN with $V$ as input. 

DFSMN is a sequential model shown in Figure 2. 
DFSMN linearly transforms the input $V$ to get $h_{t}^{1}$ and the operations of the $l$th DFSMN component take the following form: 

\vspace{-0.2cm}
\begin{align}
   \widetilde{p}_{t}^{l}= H\left ( \widetilde{p}_{t}^{l-1} \right )+p_{t}^{l}+\sum_{i=0}^{N_{1}^{l}}a_{i}^{l}\odot p_{t-s_{1}\ast i}^{l}+\sum_{i=0}^{N_{2}^{l}}c_{j}^{l}\odot p_{t+s_{2}\ast j}^{l},
\end{align}
\vspace{-0.5cm}
\begin{align}
h_{t}^{l+1}=f\left ( {U}^{l} \widetilde{p}_{t}^{l}+{d}^{l}\right),
\end{align}
\vspace{-0.6cm}
\begin{align}
   p_{t}^{l}=V^{l}h_{t}^{l}+b^{l},
\end{align}
where $N_{1}^{l}$ and $N_{2}^{l}$ denote the number of backward viewing steps and forward viewing steps of the $l$th memory block, respectively. 
$H(\cdot)$ denotes the jump connection within the memory block, which can be any linear or nonlinear transformation. 
$s_{1}$ and $s_{2}$ are respectively the encoding stride factors for historical and future moments. $\odot$ denotes element-wise vector multiplication.  $V^{l}$, $b^{l}$, $U^{l}$, $d^{l}$, $a_{i}^{l}$ and $c_{i}^{l}$ represent tunable weights.

DFSMN encodes the front and back units of each hidden state jointly, which enables the network to capture the subtle variations within a speaker's voice. These variations are essential for distinguishing between speakers, as they often reflect the individual's unique speech patterns and characteristics. Additionally, DFSMN introduces skip connections to ensure the spread of information between deeper layers and alleviate the issues of gradient vanishing or explosion. 
\vspace{-0.3cm}
\begin{figure}[ht]
    \centering
    \includegraphics[width=0.8\linewidth]{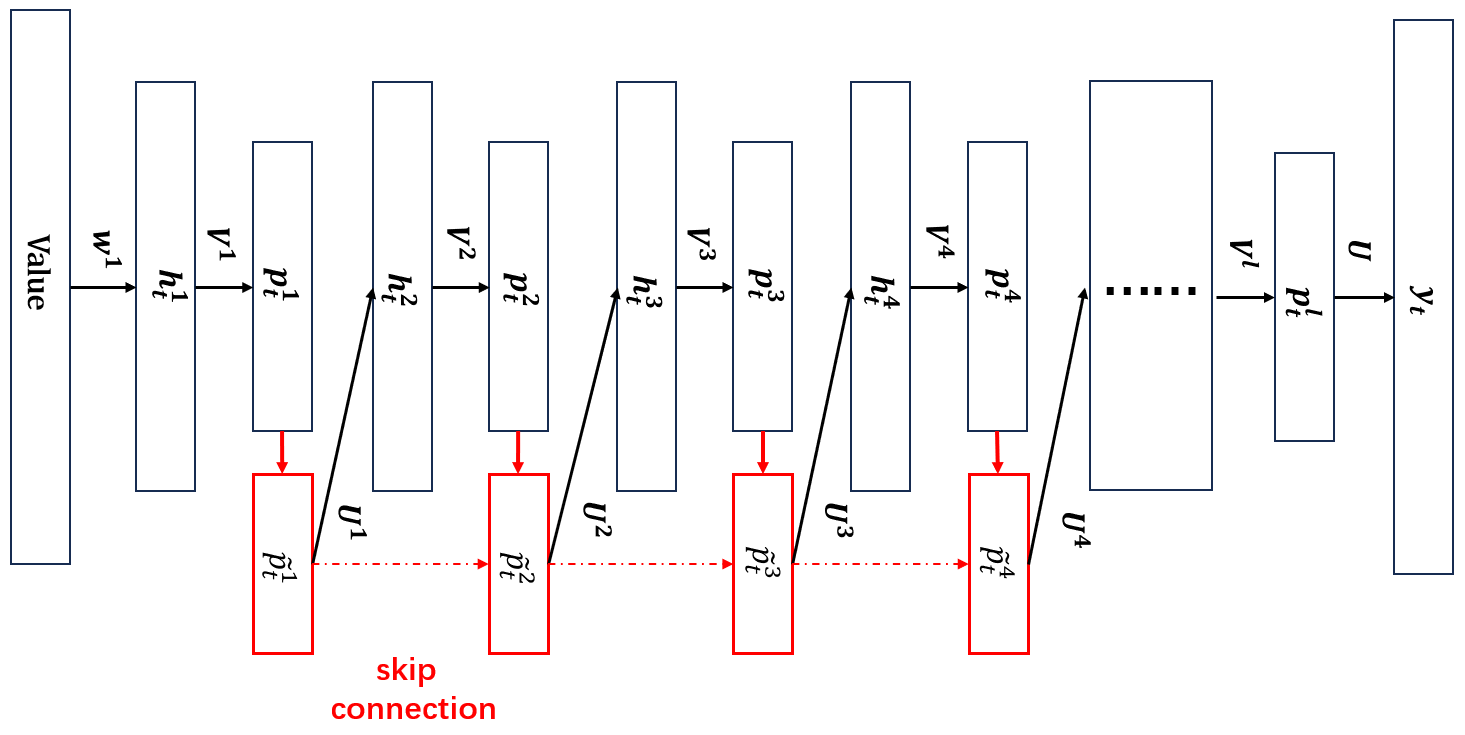}
    \caption{The architecture of DFSMN}
    \label{fig:2}
\end{figure}
\vspace{-0.3cm}
\subsection{Attentive Statistics Pooling}
The attentive statistics pooling solves the problem of ignoring the relationship between different feature frames in the traditional statistical pool by introducing attention mechanism, allowing the network to dynamically learn the importance of each feature frame and adjust the pooling results accordingly. By calculating the frame-level mean and variance based on statistical methods, the attentive statistics pooling layer averages all the frame-level features of a single utterance to form utterance-level features, thus effectively capturing the long-term changes of speech segments. Therefore, the feature mapping is finally processed through attentive statistics pooling to generate a fixed-dimensional robust speaker embedding.

\subsection{Additive Angular Margin Focal Loss (AAMF)}
In the actual speech data, we find that there are many samples that are difficult to be processed by the model, including imbalanced speech categories, significant sample noise, and the prevalence of boundary samples. These samples are called hard samples.
The purpose of the proposed AAMF is to make intra-class features closer and inter-class features farther apart by converting the traditional distance of features into angular interval, adding angular interval penalty and paying more attention to hard samples. 
Given some speech segments samples in the training set,the class number is $N$, let $y_i$ represent a category and $\theta_j$ represent the angle between the $j$th category weight and feature. We design AAMF loss function as follows:
\vspace{-0.2cm}
\begin{align}
    L_{AAMFS}=-\left(1-p_t\right)^\gamma \log \left(p_t\right),
\end{align}
with
\vspace{-0.3cm}
\begin{align}
p_t= \begin{cases}\frac{e^{s \cdot \cos (\theta_{y_i}+m)}}{e^{s \cdot \cos (\theta_{y_i}+m)}+\sum_{j=1, j \neq y_i}^N e^{s \cdot \cos \theta_j}} & \text {if } {j = y_i} \\ 1-\frac{e^{s \cdot \cos (\theta_{y_i}+m)}}{e^{s \cdot \cos (\theta_{y_i}+m)}+\sum_{j=1, j \neq y_i}^N e^{s \cdot \cos \theta_j}} & \text {otherwise, }\end{cases} 
\end{align}
where $s$ is radius of hypersphere, $m$ is an additive angular margin, $\gamma$ denotes adjustable factor, $p_t$ denotes the ease or difficulty level of samples.

When $p_t$ approaches $1$, it indicates high confidence in sample classification, and the loss function value approaches 0, reducing the model's focus on easy to classify samples. On the contrary, when $p_t$ approaches 0, it indicates that the confidence level of sample classification is low, and the values of the loss function and $\log(1-p_t)$will increase, thereby increasing the model's attention to challenging samples. This adjustment aims to reduce the mismatch between training on closed sets and evaluating on open sets, thereby improving the generalization performance of the model.
When $ p_t $ is constant, the model pays equal attention to all samples, so it no longer has the ability to distinguish hard samples. At this time, loss is what we often call AAMsoftmax [23] and the formula is as follows: 
\vspace{-0.2cm}
\begin{align}
    \mathcal{L}_{A A M S} & =- \log \frac{e^{s \cdot \cos (\theta_{y_i}+m)}}{e^{s \cdot \cos (\theta_{y_i}+m)}+\sum_{j=1, j \neq y_i}^N e^{s \cdot \cos \theta_j}}.
\end{align}

The adjustment we propose aims at reducing the mismatch between training on closed sets and evaluation on open sets, thus improving the generalization performance of the model.

\section{Experiments and Results}
\subsection{Datasets and Features}
We conducted experiments on VoxCeleb1 dataset [24] and CN-Celeb2 dataset [25]. VoxCeleb1 comprises over 100,000 speeches from 1,251 celebrities, covering 11 different speech genres and representing text-independent speech data in real-world scenarios. CN-Celeb2 includes data from Chinese speakers, aiming to establish a cross-lingual and more unconstrained testing environment. We trained on a training set containing 1,211 speakers from VoxCeleb1 and tested on both the VoxCeleb1 test set and the CN-Celeb2 test set.
For both training and testing, we randomly extracted fixed-length 3-second segments from each speech to simulate short speech segments. We used 112-dimensional filter bank features extracted using the Librosa toolkit as input to the network.\enspace These features consist of 111 mel filters and one energy dimension.\enspace The minimum cutoff frequency of the Mel filter bank was set to 20, and a pre-emphasis coefficient of 0.97 was applied. We utilized a custom window function named Povey [26]. 

\subsection{Experimental Details}
Our implementation was based on the PyTorch framework [27], utilizing the Adam optimizer with an initial learning rate of 0.0005. We applied a 25\% degradation every 2 epochs and set the weight decay to 5e-05.\enspace The batch size was set to 64, and the input consisted of 300 frames. In the testing phase, we extracted speaker embeddings from 300 frames of the speech signal. We employed three different loss functions: AM-Softmax [28], AAM-Softmax [23] and the proposed AAMF. For verification metrics, we utilized Equal Error Rate (EER) and Minimum Detection Cost Function (minDCF). All experiments were independently repeated three times to minimize the effects of random initialization.
\vspace{-0.3cm}
  \begin{table}[ht]
    \centering
    \caption{The serial or parallel  transformer structure design}
    \renewcommand{\arraystretch}{0.8}
    \begin{tabular}{ccccc}
        \toprule
        structure & layers & Hidden size & Heads  \\
        \midrule
        variant1  &\{12,6,6\} &\{2048,1024,1024\}  &\{8,4,2\}  \\
        variant2  &\{20,12\}  &\{2048,1024\}       &\{8,4\}  \\
        variant3  &\{20,6\}   &\{2048,1024\}       &\{8,4\}    \\
        variant4  &\{12,6\}   &\{2048,1024\}       &\{8,4\}    \\
        \bottomrule
    \end{tabular}
    \label{111}
\end{table}
\vspace{-0.3cm}

\subsection{Ablation Experiments}
 As shown in Figure 3, we considered the influence of the presence or absence of memory mechanisms, as well as the effects of serial or parallel configurations on the experimental results, so We validated the performance of the proposed voice encoder (Figure 3(b)) and transformer encoder (Figure 3(a)), as well as whether the parallel multi-scale VOT (Figure 3(c)) outperforms the serial configuration(Figure 3(b)). We also hypothesized that different speech segments would affect recognition results. Therefore We examined eight variations of the structure, as shown in Table 1, which were configured with different numbers of layers and multiple heads (four variations each for serial and parallel structures). 

 The results in Table 2 indicate that our proposed voice encoder outperforms the traditional transformer encoder in the speaker verification task, with a slight increase in parameters but a decrease of 19.70\% in EER and 13.93\% in minDCF. We believe that the memory mechanism compensates for the deficiency of the transformer in extracting local information, enabling it to more effectively capture both short-term and long-term relevant information. Furthermore, there is a significant difference in accuracy between the standard architecture and the parallel architecture, with the parallel architecture showing a 10.26\% decrease in EER and a slight reduction in the number of parameters. We suspect that serial encoder with different scales may disrupt the consistency of speech features. For speech input, parallel aggregation of features aligns more with human perception, enabling the acquisition of more aggregated features and finer granularity.
\vspace{-0.3cm}
\begin{figure}[ht]
    \centering
    \includegraphics[width=1\linewidth]{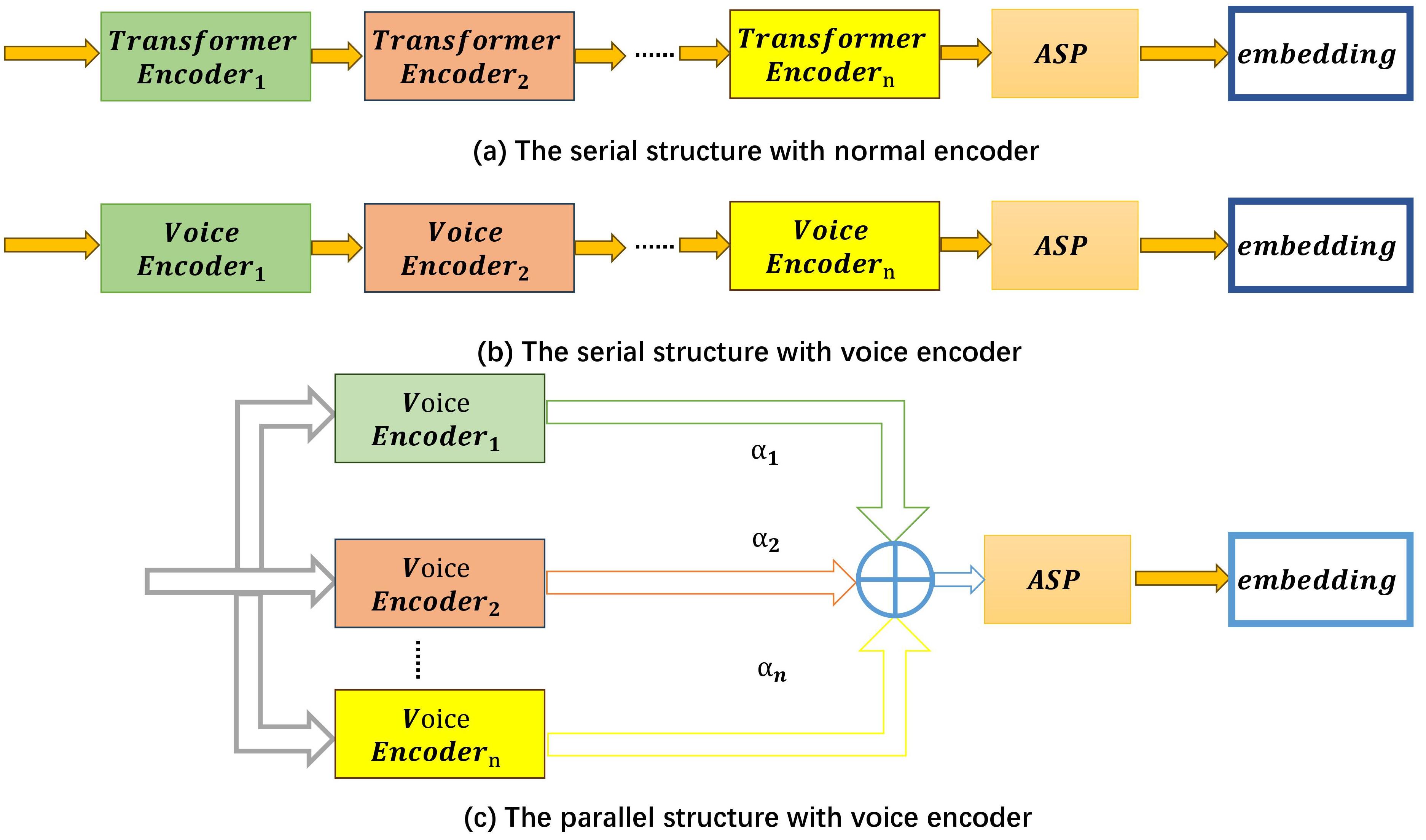}
    \caption{serial and parallel transformer structure architecture}
    \label{fig:3}
\end{figure}
\vspace{-0.3cm}

\begin{table}[ht]
    \centering
    \caption{various comparison:EER/minDCF (p=0.05) Analysis on VoxCeleb1 using different encoder, $\uparrow$ indicates the rise of indicators, $\downarrow$indicates the decline of indicators}
    \renewcommand{\arraystretch}{1}
    \setlength{\tabcolsep}{1mm}
    \label{tbl:table4}
    \begin{tabular}{cccc}
    \hline structure  &Params(MB) &EER(\%)   &minDCF\\
    \hline variant1(trans-encoder)  &23.42    &9.01    &0.4913  \\
           +serial(voice-encoder)   &25.34    &8.23    &0.4875  \\
           +parallel(voice-encoder) &25.28    &7.12    &0.4395  \\
           $\mathbf{contrast(\%)}$  &$\mathbf{7.94}\uparrow$    &$\mathbf{20.98}\downarrow$    &$\mathbf{10.54}\downarrow$  \\
    \hline variant2(trans-encoder)  &34.33    &9.14    &0.4982  \\
           +serial(voice-encoder)   &36.89    &8.22    &0.4711  \\
           +parallel(voice-encoder) &36.84    &6.86    &0.4102  \\
           $\mathbf{contrast(\%)}$  &$\mathbf{7.31}\uparrow$    &$\mathbf{24.95}\downarrow$    &$\mathbf{17.66}\downarrow$  \\
    \hline variant3(trans-encoder)  &30.38    &8.54    &0.4960  \\
           +serial(voice-encoder)   &32.46    &8.30    &0.4841  \\
           +parallel(voice-encoder) &32.41    &7.35    &0.4059  \\
           $\mathbf{contrast(\%)}$  &$\mathbf{6.68}\uparrow$    &$\mathbf{13.93}\downarrow$    &$\mathbf{18.17}\downarrow$  \\
    \hline variant4(trans-encoder)  &19.87    &8.97    &0.4767  \\
           +serial(voice-encoder)   &21.30    &7.24    &0.4576  \\
           +parallel(voice-encoder) &21.25    &7.27    &0.4321  \\
           $\mathbf{contrast(\%)}$  &$\mathbf{6.95}\uparrow$    &$\mathbf{18.95}\downarrow$    &$\mathbf{9.36}\downarrow$  \\
    \hline
    \end{tabular}
\end{table}

\begin{table*}[ht]
    \centering
    \caption{EER/minDCF (p=0.05) of other competitive methods on VoxCeleb1 and CN-Celeb2, m=0.3, $\uparrow$ indicates the rise of indicators, $\downarrow$indicates the decline of indicators. \textcolor{red}{First}, \textcolor{green}{Second}, \textcolor{orange}{Third}}
    \renewcommand{\arraystretch}{1}
    \setlength{\tabcolsep}{1.3mm}
    \label{tbl:table5}
    \begin{tabular}{ccccccc}
    \hline \multirow{2}{*}{ Method } & \multicolumn{2}{c}{ AM-Softmax } & \multicolumn{2}{c}{ AAM-Softmax } & \multicolumn{2}{c}{AAM-Softmax-Focal($\mathbf{Our}$)}\\
    \cline { 2 - 7 } & EER (\%) & minDCF & EER (\%) & minDCF & EER (\%) & minDCF\\
    \hline RawNet3(Vox1) & $8.74\pm0.17$ & $0.4942\pm0.0087$ & $8.71\pm0.22$ & $0.4882\pm0.0072$ & $8.92\pm0.14$ $\uparrow$ & $0.4963\pm0.0076$  
    $\uparrow$\\
           RawNet3(CN2) & $23.36\pm0.14$  & $0.8159\pm0.0063$ & $23.18\pm0.26$ & $0.8244\pm0.0055$ & $23.96\pm0.13$ $\uparrow$ & $0.8085\pm0.0053$ 
    $\downarrow$\\
    \hline ResNet-SE(Vox1) & $11.34\pm0.30$ & $0.5584\pm0.0121$ & $11.52\pm0.27$ & $0.5675\pm0.0108$ & $10.75\pm0.18$ $\downarrow$ & $0.5342\pm0.0096$ $\downarrow$\\
           ResNet-SE(CN2) & $23.54\pm 0.10 $  & $0.7802\pm0.0072  $ & $22.81\pm 0.14 $ &$0.8066\pm0.0069  $  & $21.79\pm 0.11 $ $\downarrow$ &$0.7923\pm0.0065  $ $\downarrow$\\
    \hline TDNN(Vox1) & $9.29\pm0.19$ & $0.4851\pm0.0098$ & $8.51\pm0.17$ & $0.4624\pm0.0102$ & $7.51\pm0.12$ $\downarrow$ & $0.4341\pm0.0091$ $\downarrow$\\
           TDNN(CN2)  & $23.04\pm 0.04 $  & $0.7752\pm0.0055  $ & $21.96\pm  0.05$ &$0.7790\pm0.0057   $ & $21.42\pm 0.12 $ $\downarrow$ &$0.7859\pm0.0060  $ $\uparrow$\\
    \hline ECAPA-TDNN(Vox1) & $\textcolor{red}{6.77\pm0.07}$ & $\textcolor{red}{0.4001\pm0.0040}$ & $\textcolor{red}{6.44\pm0.04}$ & $\textcolor{red}{0.3784\pm0.0025}$ & $\textcolor{red}{6.39\pm0.03}$ $\downarrow$ & $\textcolor{red}{0.3630\pm0.0008}$ $\downarrow$\\
           ECAPA-TDNN(CN2)  & $\textcolor{orange}{21.51\pm 0.20} $  & $\textcolor{green}{0.7540\pm0.0048 } $ & $22.03\pm 0.24 $ &$0.7775\pm0.0059  $  & $\textcolor{orange}{21.04\pm 0.18} $ $ \downarrow$ &$\textcolor{orange}{0.7526\pm0.0046}  $ $\downarrow$\\
    \hline CAM++(Vox1) & ${7.74\pm0.12}$ & $\textcolor{green}{0.4321\pm0.0077}$ & $\textcolor{green}{6.99\pm0.12}$ & $\textcolor{orange}{0.4185\pm0.0052}$ & $\textcolor{orange}{6.94\pm0.12}$ $\downarrow$ & $\textcolor{orange}{0.4152\pm0.0039}$ $\downarrow$\\
           CAM++(CN2) & $\textcolor{red}{20.63\pm  0.09}$  & $0.7761\pm0.0071  $ & $\textcolor{red}{19.52\pm 0.18} $ &$\textcolor{orange}{0.7740\pm0.0065}  $  & $\textcolor{red}{19.71\pm 0.16}$ $\downarrow$&$0.7596\pm0.0060  $ $\downarrow$\\
    \hline Transformer(Vox1) & $13.29\pm0.35$ & $0.6329\pm0.0194$ & $12.02\pm0.29$& $0.6152\pm0.0149$  & $11.69\pm0.23$ $\downarrow$ & $0.5910\pm0.0121$ $\downarrow$\\
           Transformer(CN2) & $24.89\pm 0.10 $  & $0.8203\pm0.0081  $ & $24.02\pm 0.13$ &$0.8263\pm0.0072  $  & $23.63\pm 0.05 $ $\downarrow$&$0.8156\pm0.0070  $ $\downarrow$\\
    \hline MFA-Conformer(Vox1) &  $\textcolor{orange}{7.69\pm0.25}$  & $0.4577\pm0.0049$    &$7.91\pm0.23$  & $0.4492\pm0.0012$  &$7.38\pm0.09$ $\downarrow$ &$0.4215\pm0.0021$ $\downarrow$\\
           MFA-Conformer(CN2) &  $22.12\pm0.14$  & $\textcolor{orange}{0.7660\pm0.0036}$    & $\textcolor{orange}{21.83\pm0.10}$  & $\textcolor{green}{0.7570\pm0.0024}$ & $21.16\pm0.03$ $\downarrow$ & $\textcolor{green}{0.7489\pm0.0063}$ $\downarrow$\\
    \hline VOT(ours)(Vox1) &$\textcolor{green}{7.55\pm0.10}$  &$\textcolor{orange}{0.4548\pm0.0084}$  &$\textcolor{orange}{7.58\pm0.07}$  &$\textcolor{green}{0.4179\pm0.0058}$  & $\textcolor{green}{6.86\pm0.06}$ $\downarrow$ & $\textcolor{green}{0.4102\pm0.0042}$ $\downarrow$\\
           VOT(ours)(CN2) & $\textcolor{green}{21.33  \pm0.11}  $  & $\textcolor{red}{0.7493  \pm0.0055}  $ & $\textcolor{green}{21.32  \pm0.12}  $ &$ \textcolor{red}{0.7387 \pm0.0045 } $  & $\textcolor{green}{20.38 
    \pm0.05}  $ $\downarrow$&$ \textcolor{red}{0.7352 \pm0.0048}  $ $\downarrow$\\
    \hline
    \end{tabular}
\end{table*}

\subsection{Feature Visualization}
We assessed two speech preprocessing methods and randomly selected two speaker speech features with IDs 10270 and 10271 for visual analysis, as shown in Figure 4.\enspace We compared the performance of filter bank (Fbank) and Mel frequency cepstral coefficients (MFCC) features. Our goal is to achieve high feature similarity within the speaker ID10270 class, while ensuring significant differences between speaker ID10270 and ID10271. During our experiments, we observed a notable advantage of Fbank features in speaker verification tasks, which is consistent with the clear trend observed in feature visualization.

The superiority of Fbank features in speaker verification can be attributed to their unique design, which incorporates the perceptual properties of the human ear. This design aligns more closely with the intrinsic characteristics of audio signals, enabling Fbank features to accurately capture crucial information in human speech. In contrast, the processing of MFCC features involves a linear transformation through discrete cosine transforms, potentially leading to the loss of highly nonlinear components inherent in speech signals. This renders MFCC features less effective than Fbank features for speaker verification tasks. 
\vspace{-0.5cm}
\begin{figure}[ht]
    \centering
    \includegraphics[width=0.9\linewidth]{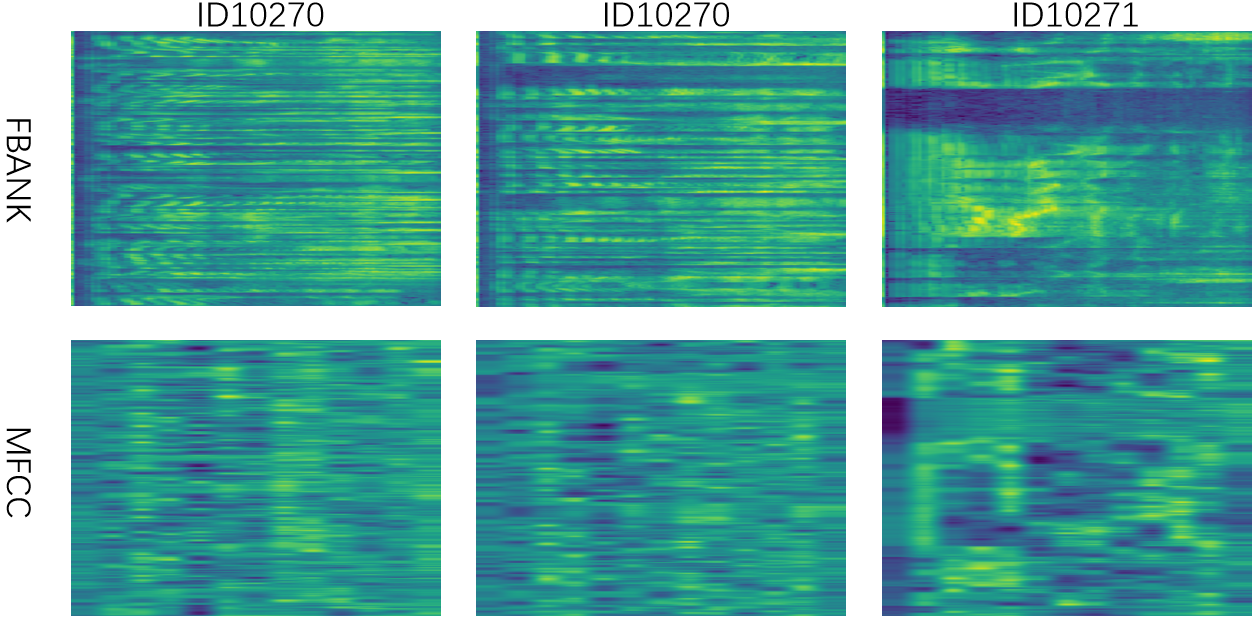}
    \caption{Comparison of features}
    \label{fig:4}
\end{figure}
\vspace{-0.5cm}

\subsection{Performance Comparison}
In this section, our model has been compared with many others using the same model weights across two distinct short speech datasets, including traditional transformers [19], classic ResNet networks [29], the currently popular ECAPA-TDNN [10], Rawnet [30], MFA-Conformer [17] and CAM++ [12]. Additionally, various network variables based on different loss functions are compared. 

The results in Table 3 demonstrate that our model outperforms all other networks except ECAPA-TDNN on the classic intra-language dataset VoxCeleb1. Trained with three different loss functions, the average EER decreases by 18.52\%, 14.55\% and 19.40\%, respectively, while the average minDCF decreases by 7.99\%, 13.43\% and 17.25\%, respectively. Surprisingly, on the challenging cross-lingual dataset CN-Celeb2, the metrics for most models undergo significant changes, indicating a lack of cross-domain adaptability. However, our model consistently maintains the best minDCF and the second-best EER. When trained with the three loss functions, the average EER decreases by 6.14\%, 3.93\% and 6.58\%, respectively, while the average minDCF decreases by 4.42\%, 6.74\% and 5.80\%, respectively. Our proposed AAMF also significantly reduces EER and minDCF of two datasets under text-independent conditions. 

We firmly believe that VOT learns commendable features, albeit being slightly inferior to the ECAPA-TDNN model. And the testing on the CN-Celeb2 dataset reveals that VOT exhibits superior stability and generalizability compared to other models, particularly in processing cross-lingual and short speech, which will be further explored in our future work. 

\section{Conclusions}
In this paper, we propose a model named VOT that incorporates memory mechanism into the attention mechanism for speaker verification. Our extensive experiments demonstrate the superiority of the memory mechanism in improving feature granularity and local modeling. The VOT model, with parallel multi-scale structure, shows improved performance compared to serial structure while reducing the parameter count. We also elucidate the suitability of deep learning approaches with Fbank features for speaker verification tasks. Additionally, our proposed loss function AAMF and the VOT model achieve excellent performance even under text-independent, short-speech, cross-lingual and unconstrained conditions. Finally, we release a PyTorch trainer for speaker verification tasks to facilitate further research in this field. 
        
\normalem
\bibliographystyle{IEEEtran}
\bibliography{mybib}
\nocite{lukic2017learning}
\nocite{liu2018speaker}
\nocite{zhu2018self}
\nocite{ioffe2006probabilistic}
\nocite{5545402}
\nocite{7078610}
\nocite{9142088}
\nocite{21701}
\nocite{8461375}
\nocite{desplanques2020ecapa}
\nocite{hu2018squeeze}
\nocite{wang2023cam++}
\nocite{thienpondt2021integrating}
\nocite{lecun1998gradient}
\nocite{zhou2019cnn}
\nocite{li2023convolution}
\nocite{zhang2022mfa}
\nocite{dosovitskiy2020image}
\nocite{vaswani2017attention}
\nocite{8461404}
\nocite{okabe2018attentive}
\nocite{lin2017focal}
\nocite{deng2019arcface}
\nocite{nagrani2017voxceleb}
\nocite{9054017}
\nocite{2012The}
\nocite{chung2020defence}
\nocite{wang2018additive}
\nocite{he2016deep}
\nocite{jung2019rawnet}

\end{document}